\documentclass[review]{elsarticle}
\usepackage{hyperref}
\journal{Cities}

\usepackage{graphics} % for pdf, bitmapped graphics files
\usepackage{graphicx} % for other environments of figures
\usepackage{subfig} % for subfigures
\usepackage{todonotes} % for dummy figures
\usepackage{upgreek} % greek font
\usepackage{array} % fancy tables
\usepackage{amsmath} % assumes amsmath package installed
\usepackage{amssymb}  % assumes amsmath package installed
\usepackage[utf8]{inputenc} % allows Spanish tildes.
\usepackage{mathtools} % loads dcases
\usepackage{multicol} % for multicols environments.

\newcommand{\bs}[1]{\boldsymbol{#1}}

\begin{document}
\begin{frontmatter}

\title{Characterization of citizens using word2vec and latent topic analysis in a large set of tweets}

\author[mainaddress]{Vladimir Vargas-Calderón}
\ead{vvargasc@unal.edu.co}

\author[secondaryaddress]{Jorge E. Camargo\corref{mycorrespondingauthor}}
\cortext[mycorrespondingauthor]{Corresponding author. System Engineering Department, Fundación Universitaria Konrad Lorenz, Carrera 9 Bis No. 62 - 43, Bogotá, Colombia, Phone: +57 1 347 23 11 ext. 211, E-mail address: jorgee.camargom@konradlorenz.edu.co}
\ead{jorgee.camargom@konradlorenz.edu.co}

\address[mainaddress]{Physics Department, Universidad Nacional de Colombia, Bogotá, Colombia}
\address[secondaryaddress]{Systems Engineering Department, Fundación Universitaria Konrad Lorenz, Bogotá, Colombia}

\begin{abstract}
With the increasing use of the Internet and mobile devices, social networks are becoming the most used media to communicate citizens' ideas and thoughts. This information is very useful to identify communities with common ideas based on what they publish in the network. This paper presents a method to automatically detect city communities based on machine learning techniques applied to a set of tweets from Bogotá's citizens. An analysis was performed in a collection of 2,634,176 tweets gathered from Twitter in a period of six months. Results show that the proposed method is an interesting tool to characterize a city population based on a machine learning methods and text analytics.

\end{abstract}


\end{frontmatter}

%\linenumbers
%%%%%%%%%%%%%%%%%%%%%%%%%%%%%%%%%%%%%%%%%%%%%%%%%%%%%%%%%%%%%%%%%%%%%%%%%%%%%%%%
\section{Introduction}
Internet usage in Colombia, and in particular in Bogotá, has been increasing in the last years, not only because it constitutes the largest source of information of every kind, but also because of governmental efforts to make people from different social backgrounds recurrent Internet users \cite{tic}, with the idea of reducing the digital gap in Colombia.	For instance, from 2011 to 2014, several portions of the city have experienced an increase of up to 148.3\% in homes with Internet connection \cite{dig_divide}.

By 2011, both Facebook and Twitter were the most popular social networks in Bogotá \cite{tic}. Moreover, 65.5\% of Bogotá's Internet users used social networks by 2014 \cite{dig_divide}. This study concentrates on Twitter because it allows its users to write posts with a length ranging from 1 to 140 characters\footnote{Data was collected before the change in the maximum tweet length to 280 characters}, making Twitter a tool for microblogging, a form of communication in which users express their opinion about several topics in short posts (tweets) \cite{tw_usage}.

By only taking the users' posts and not the explicit relations between the users, our large data set is an interesting basis for testing unsupervised community detection methodologies. In fact, the only common factor that relates the users from our data set is that they are connected by geographical location. Such unsupervised community detection methodologies help to understand social phenomena that takes place in that geographical region and in a particular period of using data from any social network with no \textit{a priori} knowledge of people's relations.

With the objective of identifying the main topics treated by Bogotá's population on Twitter, as well as detecting possible communities, we collected tweets emitted from Bogotá. Then, a Word2Vec model \cite{d2v} was built in order to represent the set of tweets corresponding to each user as a vector in a vector space. The gap statistic \cite{gap} was used to estimate the number of clusters that could be formed using these vectors. Finally, a frequency distribution of words was built for each cluster, so that each cluster could be identified by its most frequent words, which ultimately characterizes a topic.

Our work acknowledges that online social networks constitute one of the main scenarios where people express their opinions, which makes them an outstanding source of information that allows the characterization of important topics for the citizenship. Therefore, we provide a robust method for studying cultural and political aspects of a society, which unveils groups of citizens that are actively concerned about particular topics that are relevant for the dynamics of a city. The main contribution of our work is the fully unsupervised and city-dependent method that we propose for discovering such groups. In turn, the identification of the main topics discussed by the citizens simplifies the task of targeting groups of people for promoting cultural, political or educational campaigns about very specific issues.

This paper is organized as follows. In Section 2 we present related work regarding methods for topic detection, models for text representation as well as a short review on community detection algorithms. Section 3 details material and methods. In Section 4 we present the theoretical aspects of the proposed method. In Section 5 the principal results are shown. We discuss these results in Section 6 with the purpose of providing some enlightenment on the way our method detected topics treated by Bogotá's population. Finally, Section 7 presents the main conclusions of the paper and future work.

\section{Related Work}

In this section we review important milestones in the areas of topic detection, opinion-mining, text representation and community detection models, all of which are pertinent to our work.

Numerous methods have been created for topic detection in texts. The most successful and acknowledged ones are Latent Dirichlet Allocation (LDA) \cite{Blei2003}, Latent Semantic Analysis (LSA) (or Latent Semantic Indexing (LSI)) \cite{Deerwester1990}, and Correlated Topic Models (CTM) \cite{Blei2006}. However, people always tune and mix these methods in larger pipelines that allow them to study a singular application in which they are interested. In particular, analyzing posts from Twitter is a challenging task because these are short texts that lack of context. Moreover, many of these short texts are not words of an official language, both because of misspelling and because of the usage of Internet slang. However, several studies have tried to adapt to these inconveniences. A model \cite{Zhou2016} based on TwitterLDA \cite{Zhao2011} (which is an adaptation of LDA to short texts such as tweets) was constructed to take advantage of Twitter as a source of real world event information. This model aimed to detect emerging events that could affect a geographical region in a city, and also quantitatively estimated the impact of such an event on the population nearby the event location. Another study \cite{Benny2015} proposed a model to detect related topics based on clustering performed over a TF-IDF representation of tweets. The clustering was done by including weights between words called Associative Gravity Force \cite{Klahold2014} (along with other measures of similarity between words, as well as ranking of words), that accounted for the frequency of pairs of words occurring together. Also, graph-based approaches have been recently proposed \cite{Hachaj2017}, in which hash-tags are used to identify communities and topics by finding frequent co-occurring hash-tags.

Other non common methods have also been used to study texts from Twitter, like \cite{Cigarr??n2016}, in which Formal Concept Analysis (FCA) \cite{Wille1992, BernhardGanter1999}, a mathematical application of lattices and ordered sets to the process of concept formation, was used as an alternative approach to topic detection. Authors use FCA because it deals with several problems that the traditional methods suffer such as the unknown number of topics, the difficulty of these methods to adapt to new topics, among others.

It is worth noting that the majority of work has been done using English text corpora with valuable results~\cite{Dashtipour2016}. However, Spanish is significantly a more inflected language than English, and this difference could pose problems. For example, supervised machine learning methods for the topic classification of annotated Spanish tweets modeled with $n$-grams have shown to be insufficient \cite{FernandezAnta2013}. Better attempts to deal with the several research branches of topic detection and opinion-mining in Spanish have taken place. For instance, in the field of opinion mining, studies like \cite{DoloresMolina-Gonzalez2015} proposed a lexicon-based model that adapts to specific domains in Spanish for polarity classification of film reviews. Also, polarity classification in Spanish tweets has been treated in \cite{Vilares:2013:SPC:2494266.2494300}, where hybrid systems that bring together knowledge from lexical, syntactic and semantic structures in the Spanish language, as well as machine learning techniques used with the bag-of-words representation, have shown improvements over the sole bag of words approach. Besides, the clever creation of a corpus called MeSiento (Spanish for ``I feel'') \cite{Montejo-Raez:2013:SKB:2487788.2487996} allowed a robust unsupervised method for polarity classification of Spanish tweets that reached accuracy levels close to the ones obtained with supervised algorithms.

Nonetheless, these methods applied to sentiment analysis tasks depend a lot on annotated dictionaries which do not contain Bogotá's jargon. As a matter of fact, few attempts have been made to make small topic-specific dictionaries such as \cite{AlvaradoValencia2016}, where the political sentiment towards Bogotá mayoral candidates for 2015 was analyzed using Twitter and a political sentiment dictionary defined in the Colombian political context. A second study briefly examined the sentiment of tweets from Bogotá with words related with health symptoms \cite{Salcedo2015}. A third study examined the results of 2015 Colombian regional elections and compared them with political ideology and Twitter activity of the candidates \cite{Correa2016}.

In the end, it is clear that an unsupervised model for text representation is needed to give robust topic-independent text representations. One of the most successful and widely used text representation models is Word2Vec, which has proven to give good results regardless of the language in opinion mining and topic detection duties. For instance, \cite{Enr??quez2016} combined Word2Vec and a bag-of-words document classifier, and showed that Word2Vec provided word embeddings that produced more stable results when doing cross-domain classification experiments. Also, since Word2Vec was first introduced, there has been some research trying to improve and fine-tune word embeddings. Such is the case of \cite{Li2016} that proposes a hybrid model between skip-gram model and continuous bag of words (CBOW) called mixed word embedding (MWE). All in all, we choose word embedding models such as Word2Vec for being able to embed semantic similarities between words in a similarity metric defined over a Euclidean vector space.

Advances in sentiment analysis have been reported in recent works such as \cite{Dashtipour2016}, where state of the art methods were surveyed and compared. Deep learning methods are being used in works such as \cite{Dashtipour2018ExploitingDL} to analyze sentiments in Persian texts. Deep convolutional neural networks have been also investigated to analyze sentiments in Twitter \cite{Jianqiang2018}. Deep learning based methods have been used to detect malicious accounts in location-based social networks \cite{Gong2018}. One recent work used a Bayesian network and fuzzy recurrent neural networks for detecting subjectivity \cite{Chaturvedi2018}.

With regard to one of the particular objectives of our work: detecting communities, several methods have been developed in the last couple of decades to solve the so-called planted $\ell$-partition model, where the structure of graphs are studied to find densely connected groups of nodes (see refs  \cite{andrea2009,Yang2016} for excellent reviews). More modern methods based on embedding communities in low-dimensional vector spaces try to solve problems such as node clustering, node classification, low-dimensional visualizations, edges prediction, among others with great success~\cite{Cavallari:2017,yeli2018}. However, we shall point out that this is a very active area of research with many facets, and as argued in \cite{rosvall2017}, community detection should not be considered as a well-defined problem, but instead, should be motivated by particular reasons. In this sense, our motivation for detecting communities is to find groups of people with a clear topic of interest, regardless of whether such groups of people follow each other on Twitter. This means that we do not know from the beginning any connection between the nodes (users), and we aim to detect communities solely based on the data that characterizes each node, i.e. the text representation of each user's tweets.

The contribution of this paper is twofold: first, we proposed a method to automatically identify digital communities of a city grouped by topic of interest, and second, we collected a set of tweets of Bogotá's citizens to illustrate the proposed method.

\section{Material and methods\label{sec:matandmeth}}

\begin{figure}[!ht]
\centering
\includegraphics[width=1.0\columnwidth, trim={1cm 3.2cm 1cm 3cm},clip]{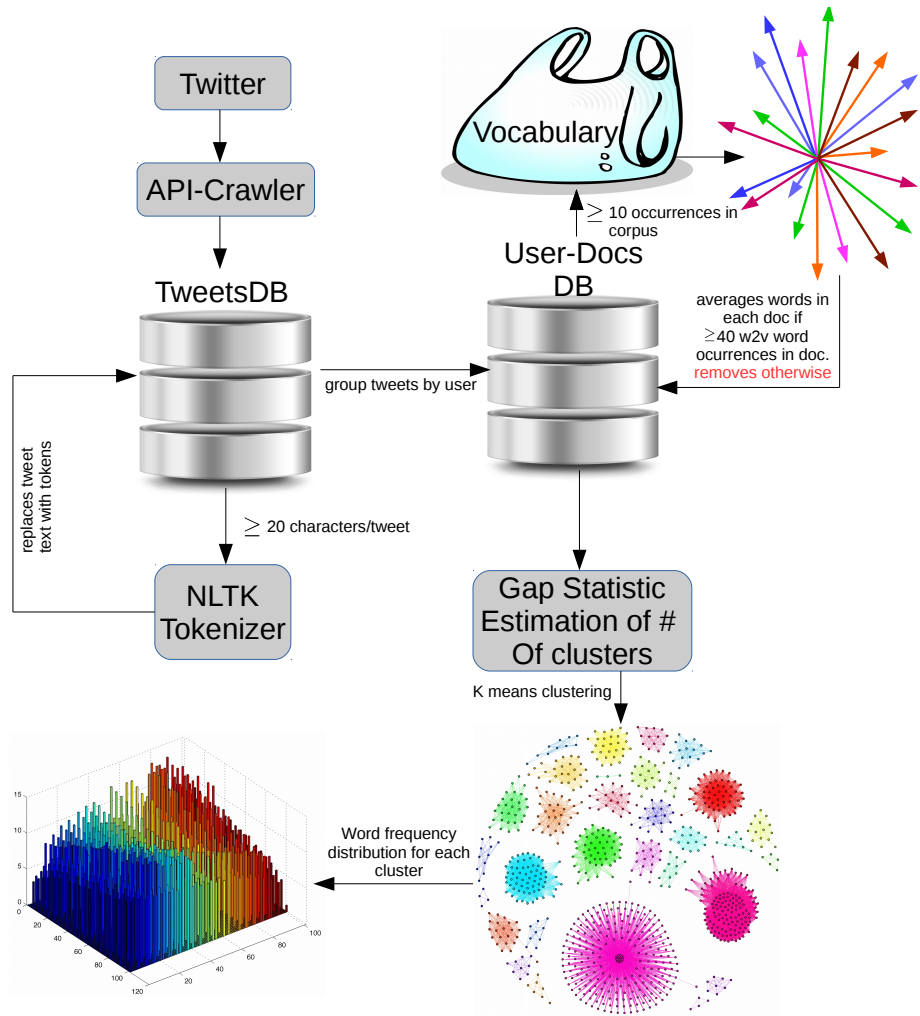}
\centering
\caption{Overview of the proposed method: (1) A component crawls a set of tweets, which are stored in a document data base; (2) The Word2Vec model is applied to this data set to build all the tweets in an embedding space; (3) A clustering analysis is performed in the embedded space to find latent topics; (4) Each user is projected in a 2D visualization in which the obtained latent topics are colored.\label{fig:flow_diag}}
\end{figure}

An overview of the proposed method is depicted in Figure \ref{fig:flow_diag}, which is inspired by the ideas found in ~\cite{Silva:2017}. We first crawl a set of tweets, which are stored in a document data base. Then, we generate vector representations of texts using Word2Vec. We selected this model because it has been the seed for all word embedding models, and it is the most widely used model, despite the existence of newer and very successful word embedding models such as fastText~\cite{fasttextzip2016,joulin2017,bojanowski2017}, BERT~\cite{bert2018}, Swivel~\cite{swivel2016} and ELMo~\cite{elmo2018}. Afterwards, clustering analysis is performed in the embedded space to find latent topics. Each user is projected in a 2D visualization in which the obtained latent topics are colored. It is of the uttermost importance to notice that we do not create a graph with explicit edges between the users, but rather let the latent topics found in each user's tweets to create implicit edges. The following sections describe in detail each of these components.

\subsection{Crawler}
A crawler component was implemented in Java using the Twitter Streaming API, which allowed to collect a set of 2,476,426 tweets of Bogotá in a period of 111 days (from August 2015 to December of 2015). The query only searched for tweets that matched with the string ``Bogotá'' in the field ``place'' of the tweet meta-data.

\subsection{Tweets data set}
\begin{figure}[!ht]
\centering
\includegraphics[width=\columnwidth]{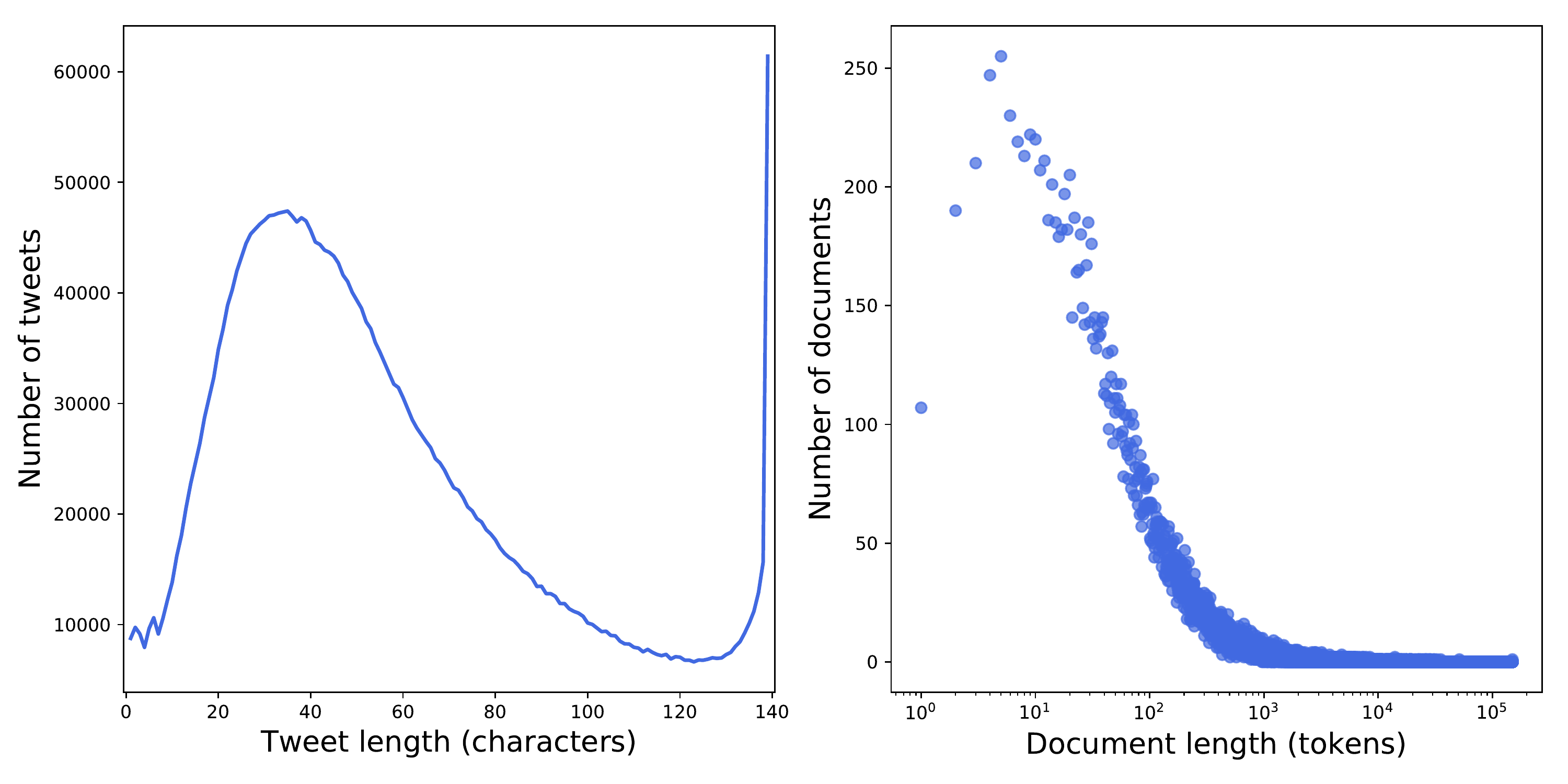}
\caption{Distribution of tweet and document length in the corpus. \label{fig:tw_dist}}
\end{figure}

Tweets corresponding to the same user form a \textit{document}. All the documents form the \textit{corpus}. The set of words composing the corpus is the \textit{vocabulary}. The distribution of tweet and document length of the corpus is shown in Figure~\ref{fig:tw_dist}. It is worth noting that most of the tweets have 20-60 characters. The average length in characters of the tweets is 55 and the average length in tokens of the documents is 639.

\subsection{Pre-processing}
We decided to discard tweets with less than 20 characters because very short texts usually lack of significant information. From the tweets with 20 or more characters, a total of 58,644 documents were created. The documents were tokenized using the NLTK's Tweet Tokenizer \cite{bird2009natural}, which allowed to preserve emoticon-like (and smileys) characters in tweets. The tokens were not stemmed because Word2Vec deals with different conjugations of words.

\subsection{Word2Vec embedding}
With these documents, a Word2Vec model with a context distance of 6 words was trained. Words that appeared at least 10 times in the corpus were selected from the vocabulary to train the model. This subset of the corpus vocabulary (or model's vocabulary) was composed of 55,168 words. Word2Vec represents each document (the average of the embedded vector words composing the document) as a vector in a vector space whose cardinality was set to 150 dimensions. 

\subsection{Documents database}
Having the vector representation in the 150 dimensional space, we discarded the ones that contained less than 40 word occurrences from the model's vocabulary. This was done with the purpose of examining documents that represented active Twitter users. A total of 30,746 documents satisfied this condition.

\subsection{Clustering of documents (users)}
In order to identify the main topics treated by Bogotá's Twitter users we used the k-means algorithm (with the Python's scikit-learn module \cite{scikit-learn}) on the vector representation of the documents. To determine the number of clusters $k$ in the cluster analysis, a gap statistic study was performed.

\subsection{Frequency distribution of words within clusters}
Once $k$ was estimated, we took the 15 most representative documents of each cluster to build a frequency distribution of words that allowed us to easily identify the topics represented in each cluster. Moreover, a tweet length distribution of the 15 documents of each cluster was also built in order to recognize which topics demanded users to write longer or shorter tweets.

\section{Theoretical framework}
This section presents the theoretical framework used in each component of the proposed method.

\subsection{Word2Vec Embedding}
\begin{figure}[!ht]
\centering
\includegraphics[width=\columnwidth]{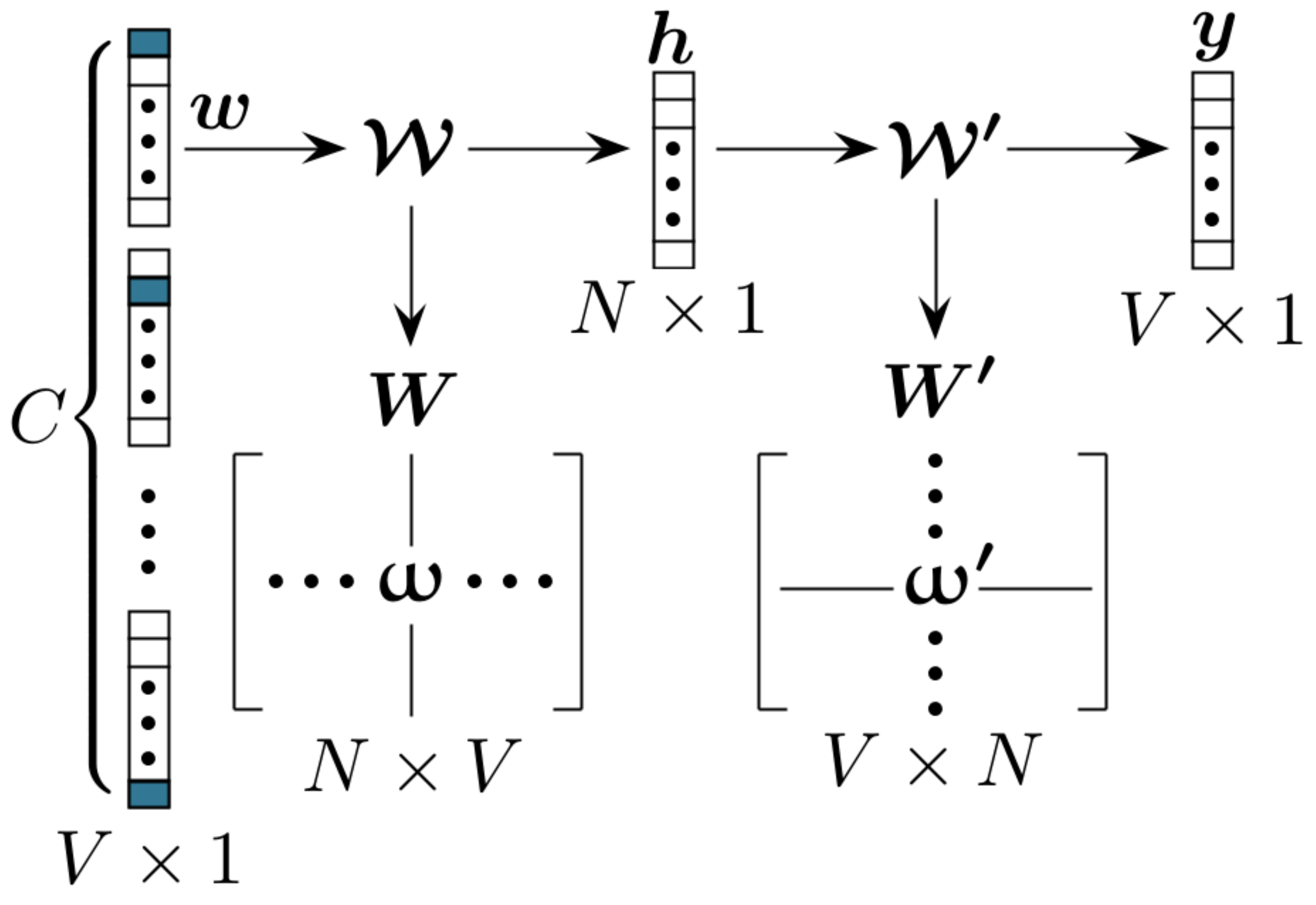}
\caption{Three-layered neural network explaining CBOW method from Word2Vec.\label{fig:nn}}
\end{figure}

Word embeddings constitute a solution to the task of numerically representing words as semantics and syntax carriers within sentences. Let $\mathcal{V}=\{w_i: i=1,2,\ldots,V\}$ be the ordered set of words, or vocabulary, contained in the corpus, where $w_i$ is the $i$-th word, and $V$ is the size of the vocabulary. Generally, words are represented as hot-vectors, which are the $V$-dimensional canonical basis vectors,
\begin{align}
\bs{w}_1 = (1,0,\ldots, 0),\ldots,\bs{w}_V = (0,0,\ldots,1),
\end{align}
where $\bs{w}_i$ is the hot-vector representation of the word $w_i$. Clearly $\langle\bs{w}_i, \bs{w}_j\rangle:=\sum_k w_{ik}w_{jk}=\delta_{ij}$, where $\delta_{ij}$ is the Kronecker's delta and $w_{ik}$ is the $k$-th component of the $i$-th word hot-vector, which shows that in this canonical basis of $\mathbb{R}^V$ there cannot be a relation between words, since they are pairwise orthonormal. Therefore, Mikolov \textit{et al.} \cite{mikolov} proposed a model to build a projector $\bs{\mathcal{W}}:\mathbb{R}^V\to\mathbb{R}^N$, where $N\ll V$, that maps hot-vectors into embedded $N$-dimensional vectors. That is, $\bs{\mathcal{W}}(\bs{w}_i)\mapsto \bs{\upomega}_i$, where $\bs{\upomega}_i$ is the embedded $N$-dimensional representation of the word $w_i$. The way of constructing $\bs{\mathcal{W}}$ is via a multi-layered neural network that can be arranged in two different ways. The first way results in the Continuous Bag of Words (CBOW) model, and the second way results in the skip-gram model. This work centers in the CBOW model, in which the neural network has the job of predicting a target word given a set of words called \emph{context words}. The context words of a word are defined as the set of words that are at a distance less than or equal to $c$ from each occurrence of the word in the corpus, where $c$ is some integer that one defines. For instance, if the corpus contains food reviews, one would expect the neural network to predict ``food'' when the context words are $\{$``delicious'', ``yummy'', ``exquisite''$\}$, and do not predict ``cat''. The neural network can be depicted as in Figure~\ref{fig:nn}, where the input layer consists of $C\times (V\times 1)$ neurons ($C$ is the number of words in the set of context words), the hidden layer of $N$ neurons, and the output layer of $V$ neurons.

To see how $\bs{\mathcal{W}}$ is constructed, consider the simplified problem of a one word context. In this case, the input layer receives a word hot-vector $\bs{w}_i$, and acts linearly on it with a $N\times V$ matrix $\bs{W}$, whose components $W_{ji}$ are the weights between the $i$-th neuron of the input layer, and the $j$-th neuron of the hidden layer.  This layer can be represented by an $N$-dimensional vector $\bs{h}$ whose components are the sum of the weights received by each neuron, i.e. $h_k = (\bs{W}\bs{w}_i)_k = W_{ki}$. Here, $\bs{W}$ can be identified as the matrix form of the projector $\bs{\mathcal{W}}$, where the $i$-th column of $\bs{W}$ is $\bs{\upomega}_i$. In a similar fashion, $\bs{W}'$ is a $V\times N$ weight matrix between the hidden and the output layer. Note that both the $i$-th column of $\bs{W}$ and the $i$-th row of $\bs{W}'$ are $N$-dimensional vector representations of the word $w_i$. Moreover, since $h_k = W_{ki}$, then $\bs{h} = \bs{\upomega}_i$. This allows $\langle \bs{\upomega}'_j, \bs{h}\rangle =\langle \bs{\upomega}'_j, \bs{\upomega}_i\rangle $ to be a score of similarity between the different vector representations of the words $w_i$ and $w_j$. These scores allow the definition of a softmax multinomial distribution,
\begin{align}
y_j := P(w_j \vert w_i) = \dfrac{\exp(\langle \bs{\upomega}'_j, \bs{\upomega}_i\rangle)}{\sum_{k=1}^V \exp(\langle \bs{\upomega}'_k, \bs{\upomega}_i\rangle)},\label{eq:multi}
\end{align}
where $y_j$ is the output of the $j$-th neuron in the output layer. Eq.~\eqref{eq:multi} is the estimated probability of $w_i$ being a context word of $w_j$. To learn the correct probabilities, a loss function $E$ is defined as
\begin{align}
E := -\log P(w_j \vert w_i) = \log\sum_{k=1}^V \exp(\langle \bs{\upomega}'_k, \bs{\upomega}_i\rangle) - \langle \bs{\upomega}'_j, \bs{\upomega}_i\rangle.\label{eq:loss}
\end{align}

Since the goal is to minimize $E$, then $\langle \bs{\upomega}'_j, \bs{\upomega}_i\rangle$ must become larger in the learning process, meaning that words that share a context must have similar vector representations, and words that do not share contexts must have dissimilar vector representations. Notice that computing $E$ is time-consuming, since normally $V$ is a big number ($\sim10^{4}$). 

Furthermore, when $C>1$, $\bs{h}$ is defined as the average vector of the context words' vector representations. Here the loss function is $E:= -\log P(w_j|\mathcal{C})$, where $\mathcal{C}$ is a set of $C$ words chosen from $\mathcal{V}$. This probability can be expressed as in Eq.~\eqref{eq:multi} by changing $\bs{\upomega}_i$ by $\bs{h}$. Notice that there are $\binom{V}{C}$ ways of picking contexts, and the number of calculations increase dramatically. To reduce this number, negative sampling \cite{mikolov} is used. Let $\tilde{P}(w)$ be the term frequency distribution of words. Let $\tilde{Q}(w)$ be a noise distribution built by taking the probability $P(w)$ of each word and raising them to the power of 3/4 (this allows less frequent words to have more probability of being drawn from $\tilde{Q}$~\cite{goldberg2014word2vec}), and then renormalizing. The problem of minimizing $E$ is converted into a binary classification problem as follows. Given a context $\mathcal{C}$, we pick a word $w$ from $\mathcal{V}$ such that $\mathcal{C}$ are context words for $w$. This pair $(w,\mathcal{C})$ is called a true sample and is labeled by $d=1$. From the noise distribution $\tilde{Q}$, $K$ words $\{w^\text{n}_1,\ldots,w^\text{n}_K\}$ are drawn. The pairs $(w^\text{n}_k,\mathcal{C})$ are negative samples and are labeled by $d=0$. Therefore, the binary classification problem can be stated as maximizing the joint probability of $(d,w)$, where $w$ can refer to a word both from the true sample or from the negative samples \cite{dyer2014notes}:
\begin{align}
P(d,w\vert \mathcal{C}) = \begin{dcases}
\dfrac{K}{1+K}\tilde{Q}(w)	&	\text{if }d=0,\\
\dfrac{1}{1+K}\tilde{P}(w\vert\mathcal{C})	&	\text{if }d=1.
\end{dcases}\label{eq:joint_prob}
\end{align}

Since the actual empirical distribution $\tilde{P}(w\vert\mathcal{C})$ is not known, the one defined in Eq.~\eqref{eq:multi} is used. Now, to reduce the calculations, under some approximations \cite{dyer2014notes}, it is possible to write Eq.~\eqref{eq:joint_prob} as two simple equations:
\begin{align}
P(d=0\vert w,\mathcal{C}) &= \frac{1}{\exp(\langle\bs{\upomega}, \bs{h}\rangle) + 1}, \label{eq:d=0}\\
P(d=1\vert w,\mathcal{C}) &= \frac{\exp(\langle\bs{\upomega}, \bs{h}\rangle)}{\exp(\langle\bs{\upomega}, \bs{h}\rangle) + 1}.\label{eq:d=1}
\end{align}
Note that these are sigmoid functions $\sigma$, of the same argument, except for a sign. From Eq.~\eqref{eq:d=0} it can be seen that if $\bs{\upomega}$ and $\bs{h}$ are dissimilar (similar) then the mentioned probability will be close to 1 (0). Similarly, in Eq.~\eqref{eq:d=1}, if $\bs{\upomega}$ and $\bs{h}$ are similar (dissimilar), the probability will also be close to 1 (0). In this case, it can be shown that the error function takes the form
\begin{align}
E = \sum_{k=1}^K\log \sigma(\langle\bs{\upomega}^\text{n}_k, \bs{h}\rangle) - \log \sigma(\langle\bs{\upomega}, \bs{h}\rangle),
\end{align}

for a target word $w$. To see negative sampling in the context of the skip-gram model (also used instead of CBOW), see \cite{goldberg2014word2vec}.

\subsection{Gap statistic}

The gap statistic allows the estimation of the number of clusters in a data set \cite{gap}. Consider a data set $\{\bs{x}_1,\ldots,\bs{x}_N\}$ of $N$ samples and $M$ features. Let $C_r$ be the $r$-th cluster found with K-Means out of $k$ clusters (other clustering algorithms can be implemented), containing $n_r=\vert C_r\vert$ samples. The within-cluster dispersion is defined as
\begin{align}
D_r := \frac{1}{2n_r} \sum_{\bs{x}_i, \bs{x}_j\in C_r} \vert\vert\bs{x}_i-\bs{x}_j\vert\vert^2,
\end{align}
which is similar to the variance except for a factor of $1/n_r$. The total within-cluster dispersion for $k$ clusters is 
\begin{align}
W_k := \sum_{r=1}^k D_r.
\end{align}

$W_k$ is normally used to estimate the ``correct'' number of clusters $K$ via the elbow method. On the other hand, the gap statistic uses null reference distributions. These are each constructed by finding the range in the $M$-dimensional space of the samples and generating $N$ data points with a uniform distribution. Next, $k$ clusters are computed for each reference distribution. If there are $B$ such distributions, then the gap statistic is defined as
\begin{align}
\operatorname{Gap}_B(k) = \frac{1}{B}\sum_{b=1}^B \log W^{(b)}_k -\log W_k,
\end{align}
where $W_k^{(b)}$ is the total within-cluster dispersion of the $b$-th reference distribution for $k$ clusters. It can be argued \cite{gap} that $\operatorname{Gap}_B(k)$ reaches a maximum for $k=K$ when the cluster centroids are aligned in an equally spaced fashion. Also, the uncertainty of $\operatorname{Gap}_B(k)$ can be estimated to be
\begin{align}
s_k = \sqrt{1+\frac{1}{B}}\operatorname{sd}(k),
\end{align}
where $\operatorname{sd}(k)$ is the standard deviation of the logarithm of the $B$ reference distributions' total within-cluster dispersion. Thus, with a 1-sigma certainty, $K$ is the value of $k$ for which $\operatorname{Gap}_B(k)\pm s_k$ reaches its maximum.

\section{Results}

\begin{figure}[!ht]
\centering
\includegraphics[width=\columnwidth]{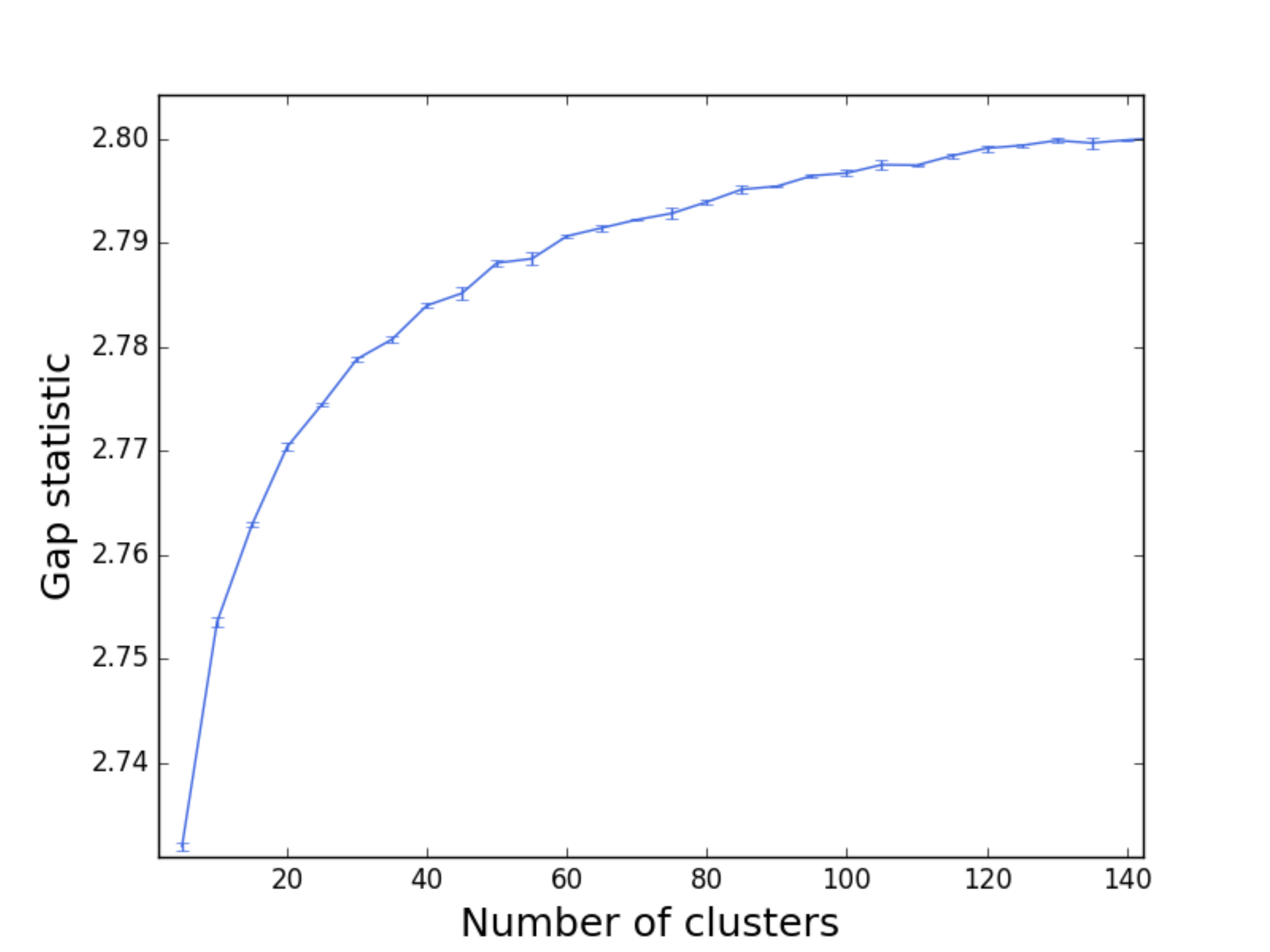}
\caption{Gap statistic for several number of clusters. \label{fig:gap}}
\end{figure}

Once the Word2Vec model was trained, the documents were represented by the average of its Word2Vec representations. The gap statistic was computed for several quantity of clusters $k=5,10,\ldots,140$, resulting in Figure~\ref{fig:gap}. The curve shown in this figure resembles a $\log$ curve. This happens because the Word2Vec model uses all the dimensions to represent the similarity relations between word vectors. Since texts are normally rich in words, it is expected that no clear clusters are formed, and therefore the number of clusters is quite indistinguishable. This is also supported by the fact that Twitter users might tweet about different topics, making their document's vector to be assigned to clusters that group 2 or more topics. Despite these inconveniences, the gap statistic allows us to estimate the number of clusters. From the figure, it can be seen that the curve begins to flatten down, or to form an elbow, around $k=40$.

\begin{figure}[h!]
\centering
\subfloat[Clusters found by k-means using Principal Component Analysis (PCA).\label{fig:pca_representative}]{\includegraphics[width=0.8\columnwidth]{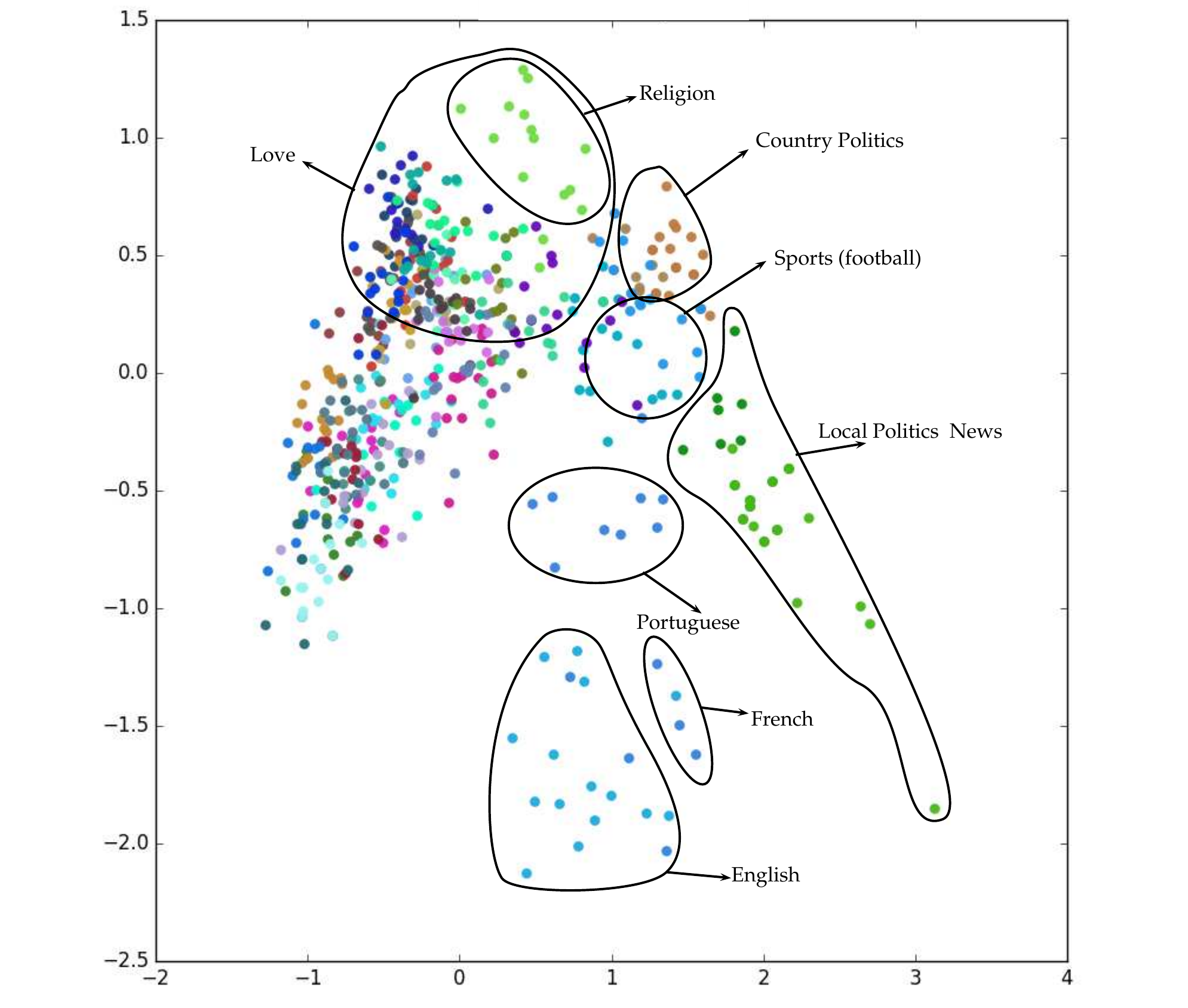}}\par\medskip
\subfloat[Clusters found by k-means using  t-distributed Stochastic Neighbor Embedding (t-SNE).\label{fig:tsne_representative}]{\includegraphics[width=0.8\columnwidth]{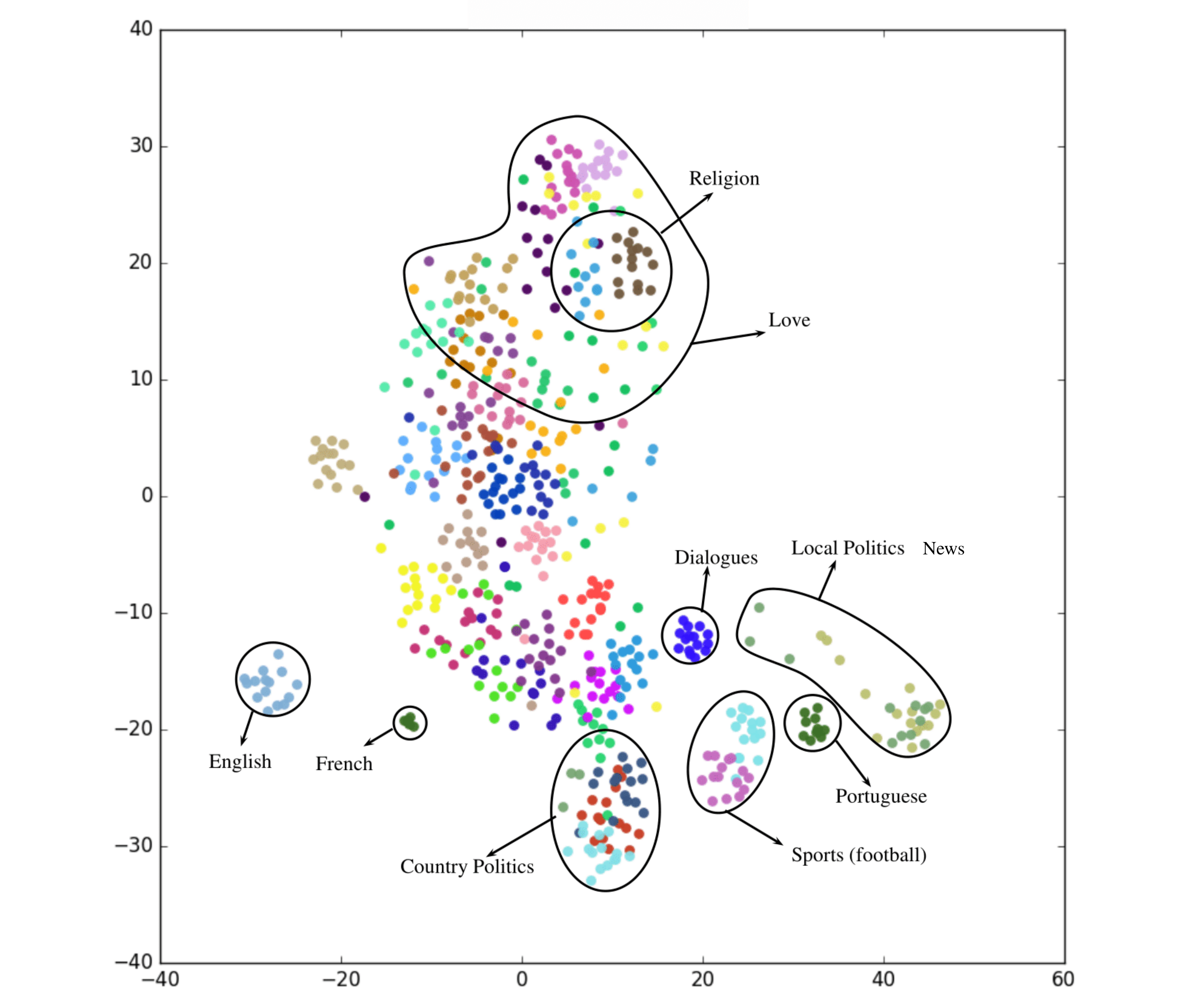}}
\caption{Annotated two-dimensional visualizations of the 15 most representative documents for each of the 40 clusters computed with K-Means. The axes of both visualizations have arbitrary units.\label{fig:visualizations}}
\end{figure}

The result of K-means clustering with 40 clusters is shown in Figure~\ref{fig:visualizations}, where the 15 most representative documents of each cluster were plotted, using PCA (Figure~\ref{fig:pca_representative}) and t-SNE (Figure~\ref{fig:tsne_representative}), using different colors for each cluster.

We established the 15 most representative documents of each cluster by sorting in ascending order the euclidean distance between each document of the cluster and its respective cluster centroid. With each set of 15 documents, a word frequency distribution was built in order to get the most frequent words of each cluster. A tweet length distribution was also built with each of these sets. The most frequent words in each cluster tell us if the topics can be easily defined. From both visualizations it is seen that there are two types of clusters: the ones corresponding to a mega-cluster, and the ones that are certainly different from the rest of the corpus, called one-topic clusters because their topic can be easily defined. The one-topic clusters are tagged in the figures with their respective topic (except from ``Love'' and ``Religion'', which can be found within the mega-cluster). It can be seen that PCA makes a sparse visualization of the documents corresponding to one-topic clusters, while t-SNE groups them and plots them apart from the mega-cluster. It should be noted that the one-topic clusters can be used to treat documents as nodes in a graph, whose edges connect nodes from the same one-topic clusters. This graph could then be used by a semi-supervised community detection method such as~\cite{mirabelli2018} in order to label the remaining nodes from the mega-cluster.

The most frequent words for each cluster are presented in Table~\ref{tab:table}. It is noteworthy that even though clusters 25 and 30 are related to the same topic, documents in cluster 30 are more news-like written, while 25 contained documents full of personal opinions about Country Politics. The same relation occurs between clusters 34 and 39. Cluster 34 is full of documents with well-written tweets that try to inform the situation of a particular football team, or match, whereas cluster 39 contains documents that express feelings about specific teams or football events. It is important to point out that cluster 2 was filled with dialogue-like written tweets, i.e. tweets that have conversations in them. In the case of clusters 22 and 24, corresponding to news, it can be identified that the combination of words used in these types of reporting tweets is very peculiar of those clusters. This is clear because even though the topic news covers plenty of topics, the way the tweets are written is perceived by Word2Vec.
\begin{table*}
\centering
\caption{The most frequent words of each cluster. Each cluster is represented by 3 cells: the first one tells the number of the cluster (\#) and the topic (in parentheses, if the topic is easily identified); the second one tells the average number of characters per tweet of that cluster; and the third one, tells the 10 most common words of the cluster (with their respective translations in English in parentheses) for the 15 clusters with the larger tweet length average, and 6 most common words for the rest of the clusters. \label{tab:table}}
\resizebox{\textwidth}{!}{%
\begin{tabular}{|c|c|c|c|c|c|c|c|c|c|}
\hline
\#22 (News) & 104 & \#30 (Country Politics) & 87 & \#34 (Sports)	&	87 	&	\#24 (News)	&	86	&	\#25	(Country Politics)	& 85\\
\hline
\multicolumn{2}{|c|}{hoy (today)} & \multicolumn{2}{|c|}{Santos (Colombia's president)}& \multicolumn{2}{|c|}{gol (goal)}& \multicolumn{2}{|c|}{via (way)}& \multicolumn{2}{|c|}{Santos (Colombia's president)}\\
\multicolumn{2}{|c|}{Tunja (a city)} & \multicolumn{2}{|c|}{FARC (Colombian guerrilla)}& \multicolumn{2}{|c|}{partido (match)}& \multicolumn{2}{|c|}{departamento (state)}& \multicolumn{2}{|c|}{FARC (Colombian guerrilla)}\\
\multicolumn{2}{|c|}{Bogotá} & \multicolumn{2}{|c|}{Colombia}& \multicolumn{2}{|c|}{city (Manchester City)}& \multicolumn{2}{|c|}{sector}& \multicolumn{2}{|c|}{paz (peace)}\\
\multicolumn{2}{|c|}{Boyacá (Colombian state)} & \multicolumn{2}{|c|}{paz (peace)}& \multicolumn{2}{|c|}{madrid (Real Madrid)}& \multicolumn{2}{|c|}{tránsito (transit)}& \multicolumn{2}{|c|}{Maduro (Venezuela's president)}\\
\multicolumn{2}{|c|}{Santa Marta (a city)} & \multicolumn{2}{|c|}{Maduro (Venezuela's president)}& \multicolumn{2}{|c|}{Santa Fé (football team)}& \multicolumn{2}{|c|}{accidente (accident)}& \multicolumn{2}{|c|}{país (country)}\\
\multicolumn{2}{|c|}{Colombia} & \multicolumn{2}{|c|}{Colombia}& \multicolumn{2}{|c|}{junior (football team)}& \multicolumn{2}{|c|}{Bogotá}& \multicolumn{2}{|c|}{país (country)}\\
\multicolumn{2}{|c|}{gobierno (government)} & \multicolumn{2}{|c|}{gobierno (government)}& \multicolumn{2}{|c|}{gran (great)}& \multicolumn{2}{|c|}{Cundinamarca (Colombian State)}& \multicolumn{2}{|c|}{terroristas (terrorists)}\\
\multicolumn{2}{|c|}{nacional (national)} & \multicolumn{2}{|c|}{justicia (justice)}& \multicolumn{2}{|c|}{mejor (better)}& \multicolumn{2}{|c|}{cierre (closure)}& \multicolumn{2}{|c|}{colombianos (Colombians)}\\
\multicolumn{2}{|c|}{vehículos (vehicles)} & \multicolumn{2}{|c|}{colombianos (Colombians)}& \multicolumn{2}{|c|}{américa (football team)}& \multicolumn{2}{|c|}{total}& \multicolumn{2}{|c|}{pueblo (people)}\\
\multicolumn{2}{|c|}{personas (people)} & \multicolumn{2}{|c|}{Uribe (Colombia's ex-president)}& \multicolumn{2}{|c|}{nacional (football team)}& \multicolumn{2}{|c|}{carril (lane)}& \multicolumn{2}{|c|}{justicia (justice)}\\
\hline
\#10 (Local Politics)	&	84	&	\#8	&	76	&	\#11 (Religion)	&	73	&	\#2 (Dialogues)	&	65	&	\#16 (Love)	&	60\\
\hline
\multicolumn{2}{|c|}{Bogotá}& \multicolumn{2}{|c|}{más (more)}& \multicolumn{2}{|c|}{Dios (God)}& \multicolumn{2}{|c|}{más (more)}& \multicolumn{2}{|c|}{vida (lane)}\\
\multicolumn{2}{|c|}{Peñalosa (Bogotá's mayor)}& \multicolumn{2}{|c|}{vida (life)}& \multicolumn{2}{|c|}{Señor (Lord)}& \multicolumn{2}{|c|}{ser (to be)}& \multicolumn{2}{|c|}{día (day)}\\
\multicolumn{2}{|c|}{Colombia }& \multicolumn{2}{|c|}{siempre (always)}& \multicolumn{2}{|c|}{amor (love)}& \multicolumn{2}{|c|}{vida (life)}& \multicolumn{2}{|c|}{amor (love)}\\
\multicolumn{2}{|c|}{Santos (Colombia's president)}& \multicolumn{2}{|c|}{solo (only)}& \multicolumn{2}{|c|}{padre (father)}& \multicolumn{2}{|c|}{hijueputa (sonofabitch)}& \multicolumn{2}{|c|}{quiero (I want)}\\
\multicolumn{2}{|c|}{Petro (Bogotá's ex-mayor)}& \multicolumn{2}{|c|}{mejor (better)}& \multicolumn{2}{|c|}{vida (life)}& \multicolumn{2}{|c|}{amor (love)}& \multicolumn{2}{|c|}{Dios (God)}\\
\multicolumn{2}{|c|}{FARC (Colombian guerrilla)}& \multicolumn{2}{|c|}{día (day)}& \multicolumn{2}{|c|}{gracias (thanks)}& \multicolumn{2}{|c|}{bien (good)}& \multicolumn{2}{|c|}{siempre (always)}\\
\multicolumn{2}{|c|}{paz (peace)}& \multicolumn{2}{|c|}{nunca (never)}& \multicolumn{2}{|c|}{corazón (heart)}& \multicolumn{2}{|c|}{solo (only)}& \multicolumn{2}{|c|}{corazón (heart)}\\
\multicolumn{2}{|c|}{Uribe (Colombia's ex-president)}& \multicolumn{2}{|c|}{cosas (things)}& \multicolumn{2}{|c|}{Jesús (Jesus)}& \multicolumn{2}{|c|}{alguien (someone)}& \multicolumn{2}{|c|}{mejor (better)}\\
\multicolumn{2}{|c|}{gobierno (government)}& \multicolumn{2}{|c|}{gente (people)}& \multicolumn{2}{|c|}{misericordia (misericordy)}& \multicolumn{2}{|c|}{mamá (mom)}& \multicolumn{2}{|c|}{mierda (shit)}\\
\multicolumn{2}{|c|}{alcalde (mayor)}& \multicolumn{2}{|c|}{tiempo (time)}& \multicolumn{2}{|c|}{fuerza (strength)}& \multicolumn{2}{|c|}{hoy (today)}& \multicolumn{2}{|c|}{gracias (thanks)}\\
\hline
\#7 (Love)	&	57	&	\#26	&	51	&	\#23 (Love)	&	50	&	\#15 (Love) & 49	&	\#9 (Portuguese and French)	&	49\\
\hline
\multicolumn{2}{|c|}{más (more)}& \multicolumn{2}{|c|}{vida (life)}& \multicolumn{2}{|c|}{Dios (God)}& \multicolumn{2}{|c|}{quiero (life)}& \multicolumn{2}{|c|}{n\~ao (no)}\\
\multicolumn{2}{|c|}{vida (life)}& \multicolumn{2}{|c|}{mejor (better)}& \multicolumn{2}{|c|}{vida (life)}& \multicolumn{2}{|c|}{día (day)}& \multicolumn{2}{|c|}{est (is)}\\
\multicolumn{2}{|c|}{quiero (I want)}& \multicolumn{2}{|c|}{solo (only)}& \multicolumn{2}{|c|}{amor (love)}& \multicolumn{2}{|c|}{jajaja (laughter)}& \multicolumn{2}{|c|}{pas (not)}\\
\multicolumn{2}{|c|}{mierda (shit)}& \multicolumn{2}{|c|}{día (day)}& \multicolumn{2}{|c|}{siempre (always)}& \multicolumn{2}{|c|}{vida (life)}& \multicolumn{2}{|c|}{mais (more)}\\
\multicolumn{2}{|c|}{mejor (better)}& \multicolumn{2}{|c|}{años (years)}& \multicolumn{2}{|c|}{feliz (happy)}& \multicolumn{2}{|c|}{mejor (better)}& \multicolumn{2}{|c|}{quero (I want)}\\
\multicolumn{2}{|c|}{gente (people)}& \multicolumn{2}{|c|}{Colombia}& \multicolumn{2}{|c|}{corazón (heart)}& \multicolumn{2}{|c|}{hoy (today)}& \multicolumn{2}{|c|}{elle (he)}\\
\multicolumn{2}{|c|}{bien (good)}& \multicolumn{2}{|c|}{nunca (never)}& \multicolumn{2}{|c|}{gracias (thanks)}& \multicolumn{2}{|c|}{alguien (someone)}& \multicolumn{2}{|c|}{vou (you)}\\
\multicolumn{2}{|c|}{siempre (always)}& \multicolumn{2}{|c|}{siempre (always)}& \multicolumn{2}{|c|}{tiempo (time)}& \multicolumn{2}{|c|}{novio (boyfriend)}& \multicolumn{2}{|c|}{minha (my)}\\
\multicolumn{2}{|c|}{amor (love)}& \multicolumn{2}{|c|}{mundo (world)}& \multicolumn{2}{|c|}{nunca (never)}& \multicolumn{2}{|c|}{amo (I love)}& \multicolumn{2}{|c|}{hoje (today)}\\
\multicolumn{2}{|c|}{necesito (I need)}& \multicolumn{2}{|c|}{Bogotá}& \multicolumn{2}{|c|}{personas (people)}& \multicolumn{2}{|c|}{bien (good)}& \multicolumn{2}{|c|}{melhor (better)}\\
\hline
\#39 (Sports Football)	&	48	&\#5 (Love)	&	44	&\#31	&	43	&\#1	&	42	&\#29	&	42	\\
\hline
\multicolumn{2}{|c|}{partido (match)}&\multicolumn{2}{|c|}{amor (love)}&\multicolumn{2}{|c|}{días (days)}&\multicolumn{2}{|c|}{quiero (I want)}&\multicolumn{2}{|c|}{usted (formal you)}\\
\multicolumn{2}{|c|}{Santa fé (football team)}&\multicolumn{2}{|c|}{quiero (I want)}&\multicolumn{2}{|c|}{hoy (today)}&\multicolumn{2}{|c|}{jajaja (laughter)}&\multicolumn{2}{|c|}{quiero (I want)}\\
\multicolumn{2}{|c|}{gol (goal)}&\multicolumn{2}{|c|}{solo (only)}&\multicolumn{2}{|c|}{bueno (good)}&\multicolumn{2}{|c|}{usted (formal you)}&\multicolumn{2}{|c|}{vida (life)}\\
\multicolumn{2}{|c|}{hoy (today)}&\multicolumn{2}{|c|}{vida (life)}&\multicolumn{2}{|c|}{gente (people)}&\multicolumn{2}{|c|}{solo (only)}&\multicolumn{2}{|c|}{solo (only)}\\
\multicolumn{2}{|c|}{vamos (come on)}&\multicolumn{2}{|c|}{amo (I love)}&\multicolumn{2}{|c|}{vida (life)}&\multicolumn{2}{|c|}{mejor (better)}&\multicolumn{2}{|c|}{cosas (things)}\\
\multicolumn{2}{|c|}{bien (good)}&\multicolumn{2}{|c|}{dia (day)}&\multicolumn{2}{|c|}{ahora (now)}&\multicolumn{2}{|c|}{vida (life)}&\multicolumn{2}{|c|}{mejor (better)}\\
\hline
\#13	&	42	&\#38	&	41	&\#28	&	41	&\#27	&	40	&\#20	&	40	\\
\hline
\multicolumn{2}{|c|}{más (more)}&\multicolumn{2}{|c|}{más (more)}&\multicolumn{2}{|c|}{más (more)}&\multicolumn{2}{|c|}{jajaja (laughter)}&\multicolumn{2}{|c|}{amo (I love)}\\
\multicolumn{2}{|c|}{tan (so)}&\multicolumn{2}{|c|}{mejor (better)}&\multicolumn{2}{|c|}{vida (life)}&\multicolumn{2}{|c|}{más (more)}&\multicolumn{2}{|c|}{más (more)}\\
\multicolumn{2}{|c|}{vida (life)}&\multicolumn{2}{|c|}{quiero (I want)}&\multicolumn{2}{|c|}{tan (so)}&\multicolumn{2}{|c|}{fav}&\multicolumn{2}{|c|}{voy (I will)}\\
\multicolumn{2}{|c|}{hoy (today)}&\multicolumn{2}{|c|}{tan (so)}&\multicolumn{2}{|c|}{alguien (someone)}&\multicolumn{2}{|c|}{tan (so)}&\multicolumn{2}{|c|}{tan (so)}\\
\multicolumn{2}{|c|}{quiero (I want)}&\multicolumn{2}{|c|}{hoy (today)}&\multicolumn{2}{|c|}{quiero (I want)}&\multicolumn{2}{|c|}{quiero (I want)}&\multicolumn{2}{|c|}{quiero (I want)}\\
\multicolumn{2}{|c|}{solo (only)}&\multicolumn{2}{|c|}{mamá (mom)}&\multicolumn{2}{|c|}{solo (only)}&\multicolumn{2}{|c|}{vida (life)}&\multicolumn{2}{|c|}{vida (life)}\\
\hline
\#4	&	40	&\#35	&	40	&\#37	&	39	&\#14	&	39	&\#32	&	39	\\
\hline
\multicolumn{2}{|c|}{más (more)}&\multicolumn{2}{|c|}{más (more)}&\multicolumn{2}{|c|}{más (more)}&\multicolumn{2}{|c|}{más (more)}&\multicolumn{2}{|c|}{más (more)}\\
\multicolumn{2}{|c|}{vida (life)}&\multicolumn{2}{|c|}{tan (so)}&\multicolumn{2}{|c|}{usted (formal you)}&\multicolumn{2}{|c|}{quiero (I want)}&\multicolumn{2}{|c|}{ser (to be)}\\
\multicolumn{2}{|c|}{alguien (someone)}&\multicolumn{2}{|c|}{vida (life)}&\multicolumn{2}{|c|}{vida (life)}&\multicolumn{2}{|c|}{vida (life)}&\multicolumn{2}{|c|}{tan (so)}\\
\multicolumn{2}{|c|}{solo (only)}&\multicolumn{2}{|c|}{mierda (shit)}&\multicolumn{2}{|c|}{mejor (better)}&\multicolumn{2}{|c|}{tan (so)}&\multicolumn{2}{|c|}{voy (I will)}\\
\multicolumn{2}{|c|}{persona (person)}&\multicolumn{2}{|c|}{solo (only)}&\multicolumn{2}{|c|}{amor (love)}&\multicolumn{2}{|c|}{mejor (better)}&\multicolumn{2}{|c|}{solo (only)}\\
\multicolumn{2}{|c|}{ser (to be)}&\multicolumn{2}{|c|}{asi (like this/that)}&\multicolumn{2}{|c|}{Dios (God)}&\multicolumn{2}{|c|}{Dios (God)}&\multicolumn{2}{|c|}{vida (life)}\\
\hline
\#6	&	38	&\#0	&	37	&\#17	&	36	&\#18	&	36	&\#33	&	36	\\
\hline
\multicolumn{2}{|c|}{más (more)}&\multicolumn{2}{|c|}{más (more)}&\multicolumn{2}{|c|}{más (more)}&\multicolumn{2}{|c|}{más (more)}&\multicolumn{2}{|c|}{más (more)}\\
\multicolumn{2}{|c|}{vida (life)}&\multicolumn{2}{|c|}{quiero (I want)}&\multicolumn{2}{|c|}{quiero (I want)}&\multicolumn{2}{|c|}{tan (so)}&\multicolumn{2}{|c|}{amor (love)}\\
\multicolumn{2}{|c|}{quiero (I want)}&\multicolumn{2}{|c|}{tan (so)}&\multicolumn{2}{|c|}{tan (so)}&\multicolumn{2}{|c|}{mejor (better)}&\multicolumn{2}{|c|}{vida (life)}\\
\multicolumn{2}{|c|}{amor (love)}&\multicolumn{2}{|c|}{vida (life)}&\multicolumn{2}{|c|}{vida (life)}&\multicolumn{2}{|c|}{vida (life)}&\multicolumn{2}{|c|}{quiero (I want)}\\
\multicolumn{2}{|c|}{solo (only)}&\multicolumn{2}{|c|}{hoy (today)}&\multicolumn{2}{|c|}{mejor (better)}&\multicolumn{2}{|c|}{quiero (I want)}&\multicolumn{2}{|c|}{solo (only)}\\
\multicolumn{2}{|c|}{siempre (always)}&\multicolumn{2}{|c|}{día (day)}&\multicolumn{2}{|c|}{hoy (today)}&\multicolumn{2}{|c|}{jajaja (laughter)}&\multicolumn{2}{|c|}{tan (so)}\\
\hline
\#19	&	35	&\#36 (English)	&	34	&\#12	&	34	&\#3	&	33	&\#21	&	33	\\
\hline
\multicolumn{2}{|c|}{más (more)}&\multicolumn{2}{|c|}{like}&\multicolumn{2}{|c|}{más (more)}&\multicolumn{2}{|c|}{más (more)}&\multicolumn{2}{|c|}{más (more)}\\
\multicolumn{2}{|c|}{mejor (better)}&\multicolumn{2}{|c|}{love}&\multicolumn{2}{|c|}{hoy (today)}&\multicolumn{2}{|c|}{quiero (I want)}&\multicolumn{2}{|c|}{amor (love)}\\
\multicolumn{2}{|c|}{quiero (I want)}&\multicolumn{2}{|c|}{people}&\multicolumn{2}{|c|}{día (day)}&\multicolumn{2}{|c|}{siempre (always)}&\multicolumn{2}{|c|}{vida (life)}\\
\multicolumn{2}{|c|}{amor (love)}&\multicolumn{2}{|c|}{want}&\multicolumn{2}{|c|}{Dios (God)}&\multicolumn{2}{|c|}{vida (life)}&\multicolumn{2}{|c|}{solo (only)}\\
\multicolumn{2}{|c|}{vida (life)}&\multicolumn{2}{|c|}{get}&\multicolumn{2}{|c|}{vida (life)}&\multicolumn{2}{|c|}{voy (I will)}&\multicolumn{2}{|c|}{quiero (I want)}\\
\multicolumn{2}{|c|}{tan (so)}&\multicolumn{2}{|c|}{don't}&\multicolumn{2}{|c|}{solo (only)}&\multicolumn{2}{|c|}{amor (love)}&\multicolumn{2}{|c|}{ser (to be)}\\
\hline

\end{tabular}}
\end{table*}

\section{Discussion}
Remarkably, many clusters contained documents that constantly make reference to love. The similarity between these clusters can be seen because they share many common words. Also, cluster 9 is a very particular one because it encloses documents containing tweets both written in French and Portuguese.

From the average tweet length of each cluster, it can be easily seen that longer tweets generally are part of a document contained in a cluster that represents a specific topic. On the other hand, clusters with short tweet length average consist of documents with tweets that express personal experiences. From the table, it is seen that there are plenty of clusters with these sorts of documents, indicating that people tend to express their personal experiences with the same set of words and in very similar semantic expressions.

\begin{figure}[!ht]
\centering
\includegraphics[width=\columnwidth]{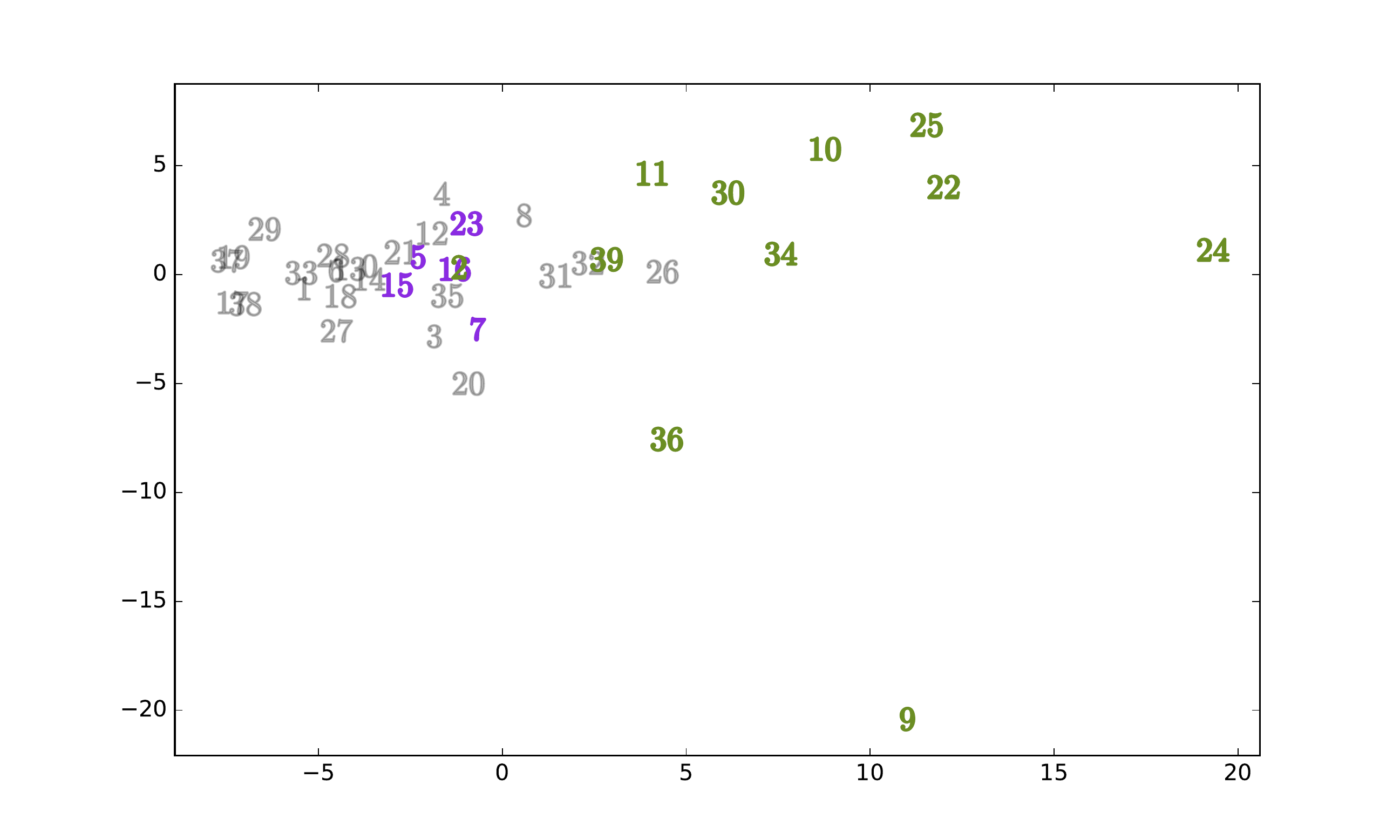}
\caption{PCA visualization of clusters' centroids. The axes have arbitrary units. Each cluster is represented by its numeric identification, as in Table~\ref{tab:table}. Green numbers show those clusters whose topic is defined. Purple numbers show the clusters whose topic was love, and the rest are the clusters whose topic is difficult to define.\label{fig:centroids}}
\end{figure}

To make Table~\ref{tab:table} easier to read, a PCA visualization of the cluster centroids is shown in Figure~\ref{fig:centroids}. In this figure, the clusters with the longer tweet length average appear to be away from the mega-cluster, comprised of all the short tweet length average clusters. This phenomenon allows us to propose that people tend to share their personal experiences in shorter tweets, while they give opinions of community important topics in longer texts.

Overlapping of document vectors in two topics means that these users belong to both communities, which is normal because typically a user is interested in more than one topic. This is known in machine learning as soft clustering, in which an object can belong to more than one cluster. In our approach this analysis was not conducted, but it would be interesting to address this aspect in future work.

\subsection{Threats to validity}

It is worth noting that our approach is not using other information available in Twitter such as user account meta-data, re-tweets, likes, URLs, images, location, geo-referenced data (when available), etc. We concentrate our analysis only in the text content. However, these complementary information could add value to better represent a user and at the same time to modify the obtained communities. Other issue in our approach is the amount of tweets used in the experimentation. We gathered data from Bogota users in a period of six months, but It is not clear what is considered enough data. However, gathering more data is a time consuming task and high demanding disk space. We considered that this issue can be addressed in future work to analyze the impact of the detected communities in a larger period of time.

\subsection{Policy implications}
It is important to remark that our work can be used by government and private entities to develop cultural, political or educational campaigns, which are the most valuable intangible fields in a society. We believe that the method proposed in this paper is especially useful in developing countries because of the gap or disconnection that exists between politicians in the government and the citizenship, because our method easily identifies the main concerns of a society. As an example, during the time we gathered the tweets, every city in Colombia was going through one of the deepest changes of the country because of the peace treaty between the Colombian government and the FARC guerrilla, which was discovered by our method as one of the main topics discussed in Bogotá. This early information could have been used by the government to explain the treaty specifically to those confused or uninformed.

\section{Conclusion and future work}
This paper presented a method to automatically identify communities using citizens data from the social network Twitter. Up to knowledge, this is the first study that analyzes Twitter data from Bogota to automatically detect communities. The use of machine learning methods such as neural networks and dimensionality reduction algorithms to detect communities is the base of the proposed method. Results show that this method can find out groups of citizens that share common topics such as politics, news, religion, sports, languages, among others. This is an interesting tool that could be used by the local government to support making decision processes in which what communities express can provide valuable information. 

As future work we want to compare our approach with the communities that can be obtained when data is modeled as a graph, which will be challenging because the obtained communities will not be necessarily comparable. We want to explore in the future the use of other data available in Twitter such as meta-data user account, re-tweets, likes, URLs, etc. It is possible that complementary information of users generates value to the concept of community. 
We also want to gather more Tweets to increase the sample data an analyze how the communities change in the time. It would be interesting to analyze how communities are influenced by social phenomena, events and period of the year.

\section*{References}

\bibliographystyle{elsarticle-num}
\bibliography{main}

\end{document}